\input{aipcheck}

\documentclass[
    ,final            
  ]
  {aipproc}

\layoutstyle{6x9}

\begin{document}

\title{Average and recommended half-life values for two neutrino double beta decay: upgrade-2013}


\author{A.S. Barabash}{
  address={Institute of Theoretical and Experimental Physics, B.\
Cheremushkinskaya 25, 117218 Moscow, Russia}
}

\begin{abstract}
 All existing positive results on two neutrino double beta decay in 
different nuclei were analyzed.  Using the procedure recommended by the 
Particle Data Group, weighted average values for half-lives of 
$^{48}$Ca, $^{76}$Ge, $^{82}$Se, $^{96}$Zr, $^{100}$Mo, $^{100}$Mo - 
$^{100}$Ru ($0^+_1$), $^{116}$Cd, $^{130}$Te, $^{136}$Xe, $^{150}$Nd, $^{150}$Nd - $^{150}$Sm 
($0^+_1$) and $^{238}$U were obtained. Existing geochemical data were 
analyzed and recommended values for half-lives of $^{128}$Te 
and $^{130}$Ba are proposed.  I recommend the use of these results as
the most currently reliable values for half-lives.
\end{abstract}

\maketitle


\section{Introduction}

  At present, the two neutrino double beta ($2\nu\beta\beta$) decay process 
has been detected in a total of 11 different nuclei. In $^{100}$Mo and 
$^{150}$Nd, this type of decay was also detected for the transition to
the $0^+$ excited state of the daughter nucleus. For the case of the 
$^{130}$Ba nucleus, evidence for the two neutrino double electron capture 
process was observed via a geochemical experiments.  All these results 
were obtained in a few tens of geochemical experiments and in $\sim$ 40 
direct (counting) experiments as well as and in one radiochemical experiment. In 
direct experiments, for some nuclei there are as many as seven independent 
positive results (e.g., $^{100}$Mo).  In some experiments, the statistical 
error does not always play the primary role in overall half-life 
uncertainties. For example, the NEMO-3 experiment with $^{100}$Mo has detected 
more than 219000 $2\nu\beta\beta$ events \cite{ARN05}, which results in a value for 
the statistical error of $\sim$ 0.2\% . At the same time, the systematic 
error in many experiments on $2\nu\beta\beta$ decay generally 
remains quite high ($\sim 10-30\%$) and very often cannot be determined 
very reliably.  As a consequence, it is frequently difficult for the 
user to select the best half-life value among the 
results. Using an averaging procedure, one can produce the most 
reliable and accurate half-life values for each isotope.

In the present work, a critical analysis of all positive experimental
results has been performed, and averaged (or recommended) values for all 
isotopes are presented.

The first time this work was done was in 2001, and the 
results were presented at MEDEX'01 \cite{BAR02}. Then revised half-life values were
presented at MEDEX'05 \cite{BAR06} and MEDEX'09 \cite{BAR09a,BAR10}. In the present paper, 
new positive results obtained since 2009 have been added and analyzed.  
\begin{table}
\caption{Present, positive $2\nu\beta\beta$ decay results. 
Here, N is the number of useful events, S/B is the signal-to-background 
ratio. $^{*)}$ For $E_{2e} > 1.2$ MeV. $^{**)}$ After correction (see text). 
$^{***)}$ For SSD mechanism. $^{****)}$ In both peaks.}
\bigskip
\label{Table1}
\begin{tabular}{|c|c|c|c|c|}
\hline
\rule[-2.5mm]{0mm}{6.5mm}
Nucleus & N & $T_{1/2}$, y & S/B & Ref., year \\
\hline
\rule[-2mm]{0mm}{6mm}
$^{48}$Ca & $\sim 100$ & $[4.3^{+2.4}_{-1.1}(stat)\pm 1.4(syst)]\cdot 10^{19}$
  & 1/5 & \cite{BAL96}, 1996 \\
 & 5 & $4.2^{+3.3}_{-1.3}\cdot 10^{19}$ & 5/0 & \cite{BRU00}, 2000 \\
& 116 & $[4.4^{+0.5}_{-0.4}(stat)\pm 0.4(syst)\cdot 10^{19}$ & 6.8 & \cite{BAR11a}, 2011 \\
\rule[-4mm]{0mm}{10mm}
 & & {\bf Average value:} $\bf 4.4^{+0.6}_{-0.5} \cdot 10^{19}$ & & \\  
          
\hline
\rule[-2mm]{0mm}{6mm}
$^{76}$Ge & $\sim 4000$ & $(0.9\pm 0.1)\cdot 10^{21}$ & $\sim 1/8$                                                        
& \cite{VAS90}, 1990 \\
& 758 & $1.1^{+0.6}_{-0.3}\cdot 10^{21}$ & $\sim 1/6$ & \cite{MIL91}, 1991 \\
& $\sim$ 330 & $0.92^{+0.07}_{-0.04}\cdot 10^{21}$ & $\sim 1.2$ & \cite{AVI91}, 1991 \\
& 132 & $1.2^{+0.2}_{-0.1}\cdot 10^{21}$ & $\sim 1.4$ & \cite{AVI94}, 1994 \\
& $\sim 3000$ & $(1.45\pm 0.15)\cdot 10^{21}$ & $\sim 1.5$ & \cite{MOR99}, 1999 
\\
& $\sim 80000$ & $[1.74\pm 0.01(stat)^{+0.18}_{-0.16}(syst)]\cdot 10^{21}$ & $\sim 1.5$ 
& \cite{HM03}, 2003 \\
& 7030 & $1.84^{+0.14}_{-0.10}\cdot 10^{21}$ & $\sim 4$ & \cite{AGO13}, 2013 \\
\rule[-4mm]{0mm}{10mm}
& & {\bf Average value:} $\bf 1.60^{+0.13}_{-0.10} \cdot 10^{21}$ & & \\  
           
\hline
\rule[-2mm]{0mm}{6mm}
& 89.6 & $1.08^{+0.26}_{-0.06}\cdot 10^{20}$ & $\sim 8$ & \cite{ELL92}, 1992 \\
$^{82}$Se & 149.1 & $[0.83 \pm 0.10(stat) \pm 0.07(syst)]\cdot 10^{20}$ & 2.3 & 
\cite{ARN98}, 1998 \\
& 2750 & $[0.96 \pm 0.03(stat) \pm 0.1(syst)]\cdot 10^{20}$ & 4 & \cite{ARN05}, 2005\\ 
& & $(1.3\pm 0.05)\cdot 10^{20}$ (geochem.) & & \cite{KIR86}, 1986 \\
\rule[-4mm]{0mm}{10mm}
& & {\bf Average value:} $\bf (0.92\pm 0.07)\cdot 10^{20}$ & & \\
 
\hline
\rule[-2mm]{0mm}{6mm}
$^{96}$Zr & 26.7 & $[2.1^{+0.8}_{-0.4}(stat) \pm 0.2(syst)]\cdot 10^{19}$ & $1.9^{*)}$ 
& \cite{ARN99}, 1999 \\
& 453 & $[2.35 \pm 0.14(stat) \pm 0.16(syst)]\cdot 10^{19}$ & 1 & \cite{ARG10}, 2010\\
& & $(3.9\pm 0.9)\cdot 10^{19}$ (geochem.)& & \cite{KAW93}, 1993 \\
& & $(0.94\pm 0.32)\cdot 10^{19}$ (geochem.)& & \cite{WIE01}, 2001 \\
\rule[-4mm]{0mm}{10mm}
& & {\bf Average value:} $\bf (2.3 \pm 0.2)\cdot 10^{19}$ & & \\

\hline
\rule[-2mm]{0mm}{6mm}
$^{100}$Mo & $\sim 500$ & $11.5^{+3.0}_{-2.0}\cdot 10^{18}$ & 1/7 & 
\cite{EJI91}, 1991 \\
& 67 & $11.6^{+3.4}_{-0.8}\cdot 10^{18}$ & 7 & \cite{ELL91}, 1991 \\
& 1433 & $[7.3 \pm 0.35(stat) \pm 0.8(syst)]\cdot 10^{18**)}$ & 3 & 
\cite{DAS95}, 1995 \\
& 175 & $7.6^{+2.2}_{-1.4}\cdot 10^{18}$ & 1/2 & \cite{ALS97}, 1997 \\
& 377 & $[6.75^{+0.37}_{-0.42}(stat) \pm 0.68(syst)]\cdot 10^{18}$ & 10 & 
\cite{DES97}, 1997 \\
& 800 & $[7.2 \pm 1.1(stat) \pm 1.8(syst)]\cdot 10^{18}$ & 1/9 & 
\cite{ASH01}, 2001 \\
& 219000 & $[7.11 \pm 0.02(stat) \pm 0.54(syst)]\cdot 10^{18***)}$ & 40 & 
\cite{ARN05}, 2005\\
& & $(2.1\pm 0.3)\cdot 10^{18}$ (geochem.)& & \cite{HID04}, 2004 \\ 
\rule[-4mm]{0mm}{10mm}
& & {\bf Average value:} $\bf (7.1\pm 0.4)\cdot 10^{18}$ & & \\

\hline
\end{tabular}
\end{table}

\addtocounter{table}{-1}
\begin{table}
\caption{continued.}
\bigskip
\begin{tabular}{|c|c|c|c|c|}

\hline
$^{100}$Mo - & $133^{****)}$ & $6.1^{+1.8}_{-1.1}\cdot 10^{20}$ & 1/7 & 
\cite{BAR95}, 1995 \\
$^{100}$Ru ($0^+_1$) &  $153^{****)}$ & $[9.3^{+2.8}_{-1.7}(stat) \pm 1.4(syst)]\cdot 
10^{20}$ & 1/4 & \cite{BAR99}, 1999 \\
 & 19.5 & $[5.9^{+1.7}_{-1.1}(stat) \pm 0.6(syst)]\cdot 10^{20}$ & $\sim 8$ & 
\cite{DEB01}, 2001 \\ 
& 35.5 & $[5.5^{+1.2}_{-0.8}(stat) \pm 0.3(syst)]\cdot 10^{20}$ & $\sim 8$ & 
\cite{KID09}, 2009 \\ 
& 37.5 & $[5.7^{+1.3}_{-0.9}(stat) \pm 0.8(syst)]\cdot 10^{20}$ & $\sim 3$ & 
\cite{ARN07}, 2007 \\ 
& $597^{****)}$ & $[6.9^{+1.0}_{-0.8}(stat) \pm 0.7(syst)]\cdot 10^{20}$ & $\sim 1/10$ & 
\cite{BEL10}, 2010 \\   
\rule[-4mm]{0mm}{10mm}
& & {\bf Average value:} $\bf 6.2^{+0.7}_{-0.5}\cdot 10^{20}$ & & \\

\hline
$^{116}$Cd& $\sim 180$ & $2.6^{+0.9}_{-0.5}\cdot 10^{19}$ & $\sim 1/4$ & 
\cite{EJI95}, 1995 \\
& 9850 & $[2.9\pm 0.06(stat)^{+0.4}_{-0.3}(syst)]\cdot 10^{19}$ & $\sim 3$ & 
\cite{DAN03}, 2003 \\
& 174.6 & $[2.9 \pm 0.3(stat) \pm 0.2(syst)]\cdot 10^{19**)}$ & 3 & 
\cite{ARN96}, 1996 \\
& 7000 & $[2.88 \pm 0.04(stat) \pm 0.16(syst)]\cdot 10^{19***)}$ & 10 & \cite{BAR11a}, 2011\\
& 4000 & $(2.5 \pm 0.5)\cdot 10^{19}$ & 5 & \cite{BAR11}, 2011\\
\rule[-4mm]{0mm}{10mm}
& & {\bf Average value:} $\bf (2.85 \pm 0.15)\cdot 10^{19}$ & & \\

\hline
\rule[-2mm]{0mm}{6mm}
$^{128}$Te& & $\sim 2.2\cdot 10^{24}$ (geochem.) & & \cite{MAN91}, 1991 \\
& & $(7.7\pm 0.4)\cdot 10^{24}$ (geochem.)& & \cite{BER93}, 1993 \\
& & $(2.41\pm 0.39)\cdot 10^{24}$ (geochem.)& & \cite{MES08}, 2008 \\
& & $(2.3\pm 0.3)\cdot 10^{24}$ (geochem.)& & \cite{THO08}, 2008 \\
\rule[-4mm]{0mm}{10mm}
& & {\bf Recommended value:} $\bf (2.0\pm 0.3)\cdot 10^{24}$ & & \\

\hline
\rule[-2mm]{0mm}{6mm}
$^{130}$Te& 260 & $[6.1 \pm 1.4(stat)^{+2.9}_{-3.5}(syst)]\cdot 10^{20}$ & 1/8 & \cite{ARN03}, 2003 \\
& 236 & $[7.0 \pm 0.9(stat) \pm 1.1(syst)]\cdot 10^{20}$ & 1/3 & \cite{ARN11}, 2011 \\
& & $\sim 8\cdot 10^{20}$ (geochem.) & & \cite{MAN91}, 1991 \\
& & $(27\pm 1)\cdot 10^{20}$ (geochem.)& & \cite{BER93}, 1993 \\
& & $(9.0\pm 1.4)\cdot 10^{20}$ (geochem.)& & \cite{MES08}, 2008 \\
& & $(8.0\pm 1.1)\cdot 10^{20}$ (geochem.)& & \cite{THO08}, 2008 \\
\rule[-4mm]{0mm}{10mm}
& & {\bf Average value:} $\bf (6.9\pm 1.3)\cdot 10^{20}$ & & \\

\hline
\rule[-2mm]{0mm}{6mm}
$^{136}$Xe & $\sim$ 50000 & $[2.30 \pm 0.02(stat) \pm 0.12(syst)]\cdot 10^{21}$ & 
10 & \cite{GAN12}, 2012 \\
& $\sim$ 19000 & $[2.172 \pm 0.017(stat) \pm 0.060(syst)]\cdot 10^{21}$ & 6 & 
\cite{ALB13}, 2013 \\

\rule[-4mm]{0mm}{10mm}
& & {\bf Average value:} $\bf(2.20\pm 0.06)\cdot 10^{21}$ & & \\

\hline
\rule[-2mm]{0mm}{6mm}
$^{150}$Nd& 23 & $[18.8^{+6.9}_{-3.9}(stat) \pm 1.9(syst)]\cdot 10^{18}$ & 
1.8 & \cite{ART95}, 1995 \\
& 414 & $[6.75^{+0.37}_{-0.42}(stat) \pm 0.68(syst)]\cdot 10^{18}$ & 6 & 
\cite{DES97}, 1997 \\
& 2018 & $[9.11^{+0.25}_{-0.22}(stat) \pm 0.63(syst)]\cdot 10^{18}$ & 2.8 & \cite{ARG09}, 2009\\
\rule[-4mm]{0mm}{10mm}
& & {\bf Average value:} $\bf(8.2\pm 0.9)\cdot 10^{18}$ & & \\

\hline
\rule[-2mm]{0mm}{6mm}
$^{150}$Nd - & $177.5^{****)}$ & $[1.33^{+0.36}_{-0.23}(stat)^{+0.27}_{-0.13}(syst)]\cdot 10^{20}$ & 
1/5 & \cite{BAR09}, 2009 \\
$^{150}$Sm ($0^+_1$) & & {\bf Average value:} $\bf 1.33^{+0.45}_{-0.26}\cdot 10^{20}$ & \\ 
 
\hline
\rule[-2mm]{0mm}{6mm}
$^{238}$U& & $\bf (2.0 \pm 0.6)\cdot 10^{21}$ (radiochem.) & & \cite{TUR91}, 1991 \\
 
\hline
\rule[-2mm]{0mm}{6mm}
$^{130}$Ba &  & $\bf 2.1^{+3.0}_{-0.8} \cdot 10^{21}$ (geochem.) & 
 & \cite{BAR96}, 1996 \\
 ECEC(2$\nu$)&  & $\bf (2.2 \pm 0.5)\cdot 10^{21}$ (geochem.) & 
 & \cite{MES01}, 2001 \\
& & $\bf (0.60 \pm 0.11)\cdot 10^{21}$ (geochem.) &
& \cite{PUJ09}, 2009 \\
\rule[-4mm]{0mm}{10mm}
& & {\bf Recommended value:} $ \bf \sim 10^{21}$ & & \\

\hline
\end{tabular}
\end{table}

\section{Present experimental data}

Experimental results on $2\nu\beta\beta$ decay in different nuclei are 
presented in Table 1.  For direct experiments, the number of useful events 
and the signal-to-background ratio are presented.

\section{Data analysis}

To obtain an average of the ensemble of available data, a standard weighted 
least-squares procedure, as recommended by the Particle Data Group 
\cite{PDG12}, was used.  The weighted average and the corresponding error 
were calculated, as follows:
\begin{equation}
\bar x\pm \delta \bar x = \sum w_ix_i/\sum w_i \pm (\sum w_i)^{-1/2} , 
\end{equation} 
where $w_i = 1/(\delta x_i)^2$.  Here, $x_i$ and $\delta x_i$ are 
the value and error reported by the i-th experiment, and 
the summations run over N experiments.  

The next step is to calculate $\chi^2 = \sum w_i(\bar x - x_i)^2$ and 
compare it with N - 1, which is the expectation value of $\chi^2$ if the 
measurements are from a Gaussian distribution.  If $\chi^2/(N - 1)$ is 
less than or equal to 1 and there are no known problems with the data, 
then one accepts the results to be sound.  If $\chi^2/(N - 1)$ is very large ($>>$ 1), 
one chooses 
not to use the average.  Alternatively, one may quote the calculated 
average while making an educated guess of the error using a conservative 
estimate designed to take into account known problems with the data.
Finally, if $\chi^2/(N - 1)$ is larger than 1 but not greatly so, it is still 
best to use the average data but to increase the quoted error, $\delta \bar x$ 
in Eq. 1, by a factor of S, defined as 
\begin{equation}
S = [\chi^2/(N - 1)]^{1/2}.
\end{equation} 
For averages, the statistical and systematic errors are treated in quadrature 
and use as a combined error $\delta x_i$. In some cases, only the results 
obtained with high enough 
signal-to-background ratio were used.

\subsection{$^{48}$Ca }    
There are three independent experiments in 
which $2\nu\beta\beta$ decay of $^{48}$Ca was observed \cite{BAL96,BRU00,BAR11a}. 
The results are in good agreement. The weighted average value is
$$
T_{1/2} = 4.4^{+0.6}_{-0.5} \cdot 10^{19} yr.
$$ 

\subsection{$^{76}$Ge } 
Let us consider the results of six 
experiments. First of all, however, a few additional comments are 
necessary:

1) Final result of the Heidelberg-Moscow Collaboration is used \cite{HM03}.  

2) In Ref. \cite{AVI91}, the value $T_{1/2} = 
0.92^{+0.07}_{-0.04}\cdot 10^{21}$ yr was presented. However, after a more 
careful analysis, this result has been changed to a value of 
$T_{1/2} = 1.2^{+0.2}_{-0.1}\cdot 10^{21}$ yr \cite{AVI94}, 
which was used in our analysis.

Finally, in calculating the average, only the results of experiments 
with signal-to-background ratios greater than 1 were used (i.e., the 
results of Refs. \cite{AVI94,MOR99,HM03,AGO13}). The weighted average value is
$$
    T_{1/2} = 1.6^{+0.13}_{-0.10} \cdot 10^{21} yr.
$$ 

\subsection{$^{82}$Se}
There are three independent counting 
experiments and many geochemical measurements $(\sim 20)$. The geochemical 
data are neither in good agreement with each other nor in good agreement 
with the data from direct measurements.  Typically, the accuracy of 
geochemical measurements is on the level of $\sim$ 10\% and
sometimes even better.  Nevertheless, the possibility of existing large 
systematic errors cannot be excluded (see discussion in Ref. \cite{MAN86}). 
Thus, to obtain a 
average half-life value for $^{82}$Se, only the results of the direct 
measurements \cite{ARN05,ARN98,ELL92} were used.  The result of Ref. 
\cite{ELL87} is the preliminary result of \cite{ELL92}; hence it has not
been used in our analysis. The result of Ref. \cite{ELL92} is presented
with very asymmetrical errors. To be more conservative, only
the top error in this case is used.
As a result, 
the weighted average value is
$$
T_{1/2} = (0.92 \pm 0.07) \cdot 10^{20} yr.
$$ 

\subsection{$^{96}$Zr} 
There are two positive geochemical results
\cite{KAW93,WIE01} and two results from the direct experiments of NEMO-2 \cite{ARN99} and 
NEMO-3 \cite{ARG10} .  Taking into account the comment in previous
section, I use the values from Refs. \cite{ARN99,ARG10} to obtain 
a present weighted half-life value for $^{96}$Zr of: 
$$
T_{1/2} = (2.3 \pm 0.2)\cdot 10^{19} yr.                    
$$ 

\subsection{$^{100}$Mo} 
Formally, there are seven positive 
results\footnote{I do not consider the result of Ref. \cite {VAS90a} 
because of a potentially high background contribution that was 
not excluded in this experiment.} from direct experiments and one 
result from a geochemical experiment. I do not consider the 
preliminary result of S. Elliott et al. \cite{ELL91} and instead use their 
final result \cite{DES97}, plus I do not use the geochemical result 
(again, see comment in section for $^{82}$Se).  Finally, in calculating the average, 
only the results of experiments with signal-to-background
ratios greater than 1 were used (i.e., the results of Refs. 
\cite{DAS95,DES97,ARN05}).  In addition, I have used the corrected 
half-life value from Ref. \cite {DAS95}. Thus the original 
result was decreased by 15\% because the calculated efficiency (by MC)
was overestimated (see Ref. \cite {VAR97}). In addition, the half-life 
value was decreased by 10\%, taking into account that for 
$^{100}$Mo, we have the Single State Dominance (SSD) 
mechanism (see discussion in \cite {ARN04,SHI06}). The following weighted average value 
for this half-life is then obtained:
$$
T_{1/2} = (7.1 \pm 0.4)\cdot 10^{18} yr .                                   
$$

\subsection{$^{100}$Mo - $^{100}$Ru ($0^+_1$; 1130.29 keV)} 
The 
transition to the $0^+$ excited state of $^{100}$Ru was detected in six 
independent experiments.  The results are in good agreement, and the 
weighted average for the half-life using results from \cite{BAR95,BAR99,KID09,ARN07,BEL10} is
$$
T_{1/2} = 6.2^{+0.7}_{-0.5}\cdot 10^{20} yr.
$$                                   
The result from \cite{DEB01} was not used here because I consider the result from \cite{KID09}
as the final result of the TUNL-ITEP experiment.

\subsection{$^{116}$Cd} 
There are five independent positive 
results that are in good agreement with each other when taking into 
account the corresponding error bars.  Again, I use here the corrected 
result for the half-life value from Ref. \cite{ARN96}.  The original 
half-life value was decreased by $\sim$ 25\% (see remark in section for $^{100}$Mo). The 
weighted average value is 
$$          
T_{1/2} = (2.85 \pm 0.15)\cdot 10^{19} yr.
$$ 

\subsection{$^{128}$Te and $^{130}$Te} 
For a long time, there were only geochemical 
data for these isotopes. Although the half-life 
ratio for these isotopes has been obtained with good accuracy $(\sim 3\%)$ 
\cite{BER93}, the absolute values for $T_{1/2}$ of each nuclei 
are different from one experiment to the next.  One group of authors 
\cite{MAN91,TAK66,TAK96} gives $T_{1/2} \approx 0.8\cdot 10^{21}$ yr  
for $^{130}$Te and $T_{1/2} \approx  2\cdot 10^{24}$ yr for $^{128}$Te, 
whereas another group \cite{KIR86,BER93} claims $T_{1/2} \approx 
(2.5-2.7)\cdot 10^{21}$ yr and  $T_{1/2} \approx 7.7\cdot 10^{24}$ yr, 
respectively. Furthermore, as a rule, experiments with young 
samples ($\sim 100$ million years) give results of the half-life value of 
$^{130}$Te in the range of $\sim (0.7-0.9)\cdot 10^{21}$ yr,
while for old samples ($> 1$ billion years) have half-life values in the
range of $\sim (2.5-2.7)\cdot 10^{21}$ yr. 


Recently it was argued that short half-lives are more likely to be correct \cite{MES08,THO08}.
Using different young mineral results, the half-life values were estimated at 
$(9.0 \pm 1.4)\cdot 10^{20}$ yr \cite{MES08} and $(8.0 \pm 1.1)\cdot 10^{20}$ yr \cite{THO08} 
for $^{130}$Te and $(2.41 \pm 0.39)\cdot 10^{24}$ y \cite{MES08} and $(2.3 \pm 0.3)\cdot 10^{24}$ yr \cite{THO08} 
for $^{128}$Te. 

The first indication of a positive result for $^{130}$Te in a direct experiment was obtained in 
\cite{ARN03}. More accurate and reliable value was obtained recently in NEMO-3 experiment \cite{ARN11}.
The results are in good agreement, and the weighted average value for half-life is
$$
T_{1/2} = (6.9 \pm 1.3)\cdot 10^{20} yr.
$$ 
Now, using very well-known ratio $T_{1/2}(^{130}{\rm Te})/T_{1/2}(^{128}{\rm Te}) =
(3.52 \pm 0.11)\cdot 10^{-4}$ \cite{BER93},
one can obtain half-life value for $^{128}$Te,
$$
T_{1/2} = (2.0 \pm 0.3)\cdot 10^{24} yr.
$$  
I recommend to use these last two results as the best present half-life values for 
$^{130}$Te and $^{128}$Te, respectively.

\subsection{$^{136}$Xe}
The half-life value was measured in two independent experiments, EXO \cite{ACK11,ALB13} 
and Kamland-Zen \cite{GAN12a,GAN12}.
To obtain average value I use most precise results from these experiments, obtained in 
\cite{GAN12,ALB13} (see Table 1). 
The weighted average value is
$$
T_{1/2} = (2.20 \pm 0.06)\cdot 10^{21} yr.
$$

\subsection{$^{150}$Nd}
This half-life value was measured in three 
independent experiments \cite{ART95,DES97,ARG09}. Using Eq. 1, and three 
existing values, one 
can obtain $T_{1/2} = (8.2 \pm 0.5)\cdot 10^{18}$ yr.  Taking into account 
the fact that $\chi^2 > 1$ and S = 1.89 (see Eq. 2) I finally obtain:
$$
T_{1/2} = (8.2 \pm 0.9)\cdot 10^{18} yr.
$$ 

\subsection{$^{150}$Nd - $^{150}$Sm ($0^+_1$; 740.4 keV)}
There is 
only one positive result from a direct (counting) experiment \cite{BAR09}:
$$          
T_{1/2} = [1.33^{+0.36}_{-0.23}(stat)^{+0.27}_{-0.13}(syst)]\cdot 10^{20} yr.
$$ 
The preliminary result of this work was published in \cite{BAR04}.

\subsection{$^{238}$U}  
There is again only one positive result, but this time from 
a radiochemical experiment \cite{TUR91}:
$$          
T_{1/2} = (2.0 \pm 0.6)\cdot 10^{21} y.
$$ 

\subsection{$^{130}$Ba (ECEC)}
For $^{130}$Ba positive results were obtained in geochemical measurements only. First positive 
result for $^{130}$Ba was mentioned in Ref. \cite{BAR96}. In this paper 
positive result was obtained for one sample of barite 
($T_{1/2} = 2.1^{+3.0}_{-0.8} \cdot 10^{21}$ yr), but for second sample only 
limit was established ($T_{1/2} > 4 \cdot 10^{21}$ yr). Then more accurate 
half-life values, $(2.2 \pm 0.5) \cdot 10^{21}$ yr \cite{MES01} and 
$(0.60 \pm 0.11)\cdot 10^{21}$ yr \cite{PUJ09}, were obtained. The results are in 
strong disagreement. One can not use usual 
average procedure in this case. One just can conclude that half-life of $^{130}$Ba is 
$\sim 10^{21}$ yr. 
To obtain more precise and correct half-life value for $^{130}$Ba 
new measurements are needed. 

\section{Conclusion}

In summary, all positive $2\nu\beta\beta$-decay results were analyzed, 
and average values for half-lives were calculated. For the cases of 
$^{128}$Te and $^{130}$Ba, so-called recommended values 
have been proposed.  I strongly recommend the use of these values as 
the most reliable presently. 






\bibliographystyle{aipproc}   

\bibliography{sample}

\IfFileExists{\jobname.bbl}{}
 {\typeout{}
  \typeout{******************************************}
  \typeout{** Please run "bibtex \jobname" to optain}
  \typeout{** the bibliography and then re-run LaTeX}
  \typeout{** twice to fix the references!}
  \typeout{******************************************}
  \typeout{}
 }

\end{document}